# Ultrafast photoinduced phase transition in the antiferromagnetic Dirac semimetal EuAgAs

Hao Liu,[1] Chen Zhang,[1] Qi-Yi Wu,[1] Yahui Jin,[2] Ziming Zhu,[2] Jiao-Jiao Song,[1] Sheng-Tao Cui,[3] Zhe Sun,[3] Honghong Wang,[4] Bo Chen,[1] Jun He,[1] Hai-Yun Liu,[5] Yu-Xia Duan,[1] Peter M. Oppeneer,[6,*] and Jian-Qiao Meng[1,†]

[1]*School of Physics, Central South University, Changsha, Hunan 410083, China*
[2]*Key Laboratory of Low-Dimensional Quantum Structures and Quantum Control of Ministry of Education, Department of Physics, Hunan Normal University, Changsha, Hunan 410081, China*
[3]*National Synchrotron Radiation Laboratory, University of Science and Technology of China, Hefei 230029, China*
[4]*Center for Quantum Materials and Superconductivity (CQMS) and Department of Physics, Sungkyunkwan University, Suwon 16419, South Korea*
[5]*Beijing Academy of Quantum Information Sciences, Beijing 100085, China*
[6]*Department of Physics and Astronomy, Uppsala University, Box 516, S-75120 Uppsala, Sweden*
(Dated: March 19, 2025)

We report the observation of a light-induced subpicosecond phase transition in the antiferromagnetic Dirac semimetal EuAgAs, achieved through ultrafast optical excitation. Using ultrafast optical spectroscopy, we probe the nonequilibrium carrier dynamics, discovering distinct fluence-dependent responses in the antiferromagnetic and paramagnetic states, and revealing a possible magnetic order-driven transition between different topological states. Our results demonstrate that EuAgAs, with its highly tunable magnetic structure, possibly offers a unique platform for exploring topological phase transitions. These results underscore the potential of ultrashort optical pulses as powerful tools for the real-time control of topological phases, opening pathways for advances in spintronics, quantum computing, and energy-efficient information technologies.

Magnetic topological materials are of considerable interest due to the coupling between magnetism and topological states, which can lead to the emergence of exotic quantum phenomena such as the quantum anomalous Hall effect [1, 2], topological magnetoelectric effects [3, 4], and the topological Hall effect [5–7]. A defining characteristic of these materials is the ability to drive topological phase transitions through the control of magnetic ordering [6–13]. This capability holds promise for developing new spintronics, quantum computing architectures, and energy-efficient information storage technologies.

Eu-based rare-earth compounds, known for their strong spin-orbit coupling and magnetic interactions mediated by localized $4f$ electrons, have gained attention as promising candidates for realizing tunable magnetic topological states. Materials such as $EuIn_2As_2$ [14, 15] and $EuCd_2As_2$ [16, 17] exhibit magnetic topological phases that respond sensitively to external perturbations such as magnetic fields and strain, enabling transitions between Dirac and Weyl semimetal states. This makes them ideal platforms for exploring the rich interplay between magnetism and topology.

In this context, EuAgAs, an antiferromagnetic (AFM) Dirac semimetal with a Néel temperature ($T_N$) of 12 K, has recently attracted attention as a promising platform for investigating the interplay of magnetism and topology [18–20]. The material's $A$-type AFM order, with ferromagnetic (FM) coupling within Eu hexagonal layers and AFM coupling between layers along the $c$ axis [20, 21], drives several interesting topological transitions. Previous studies have reported the observation of chiral anomaly-induced positive longitudinal magnetoconductivity in materials, which has been interpreted as evidence for the generation of Weyl fermion states under the influence of an applied magnetic field [20]. Despite these advances, direct experimental observation of the evolution of the topological band structure in response to changes in magnetic order remains an elusive goal.

Ultrafast laser spectroscopy offers a powerful tool for probing and manipulating topological materials [22, 23]. By rapidly perturbing the energy landscape, this technique allows for the detection of subtle changes in electronic structure and the induction of topological phase transitions. Previous studies have demonstrated the efficacy of ultrafast spectroscopy in investigating high-temperature superconductors [24–26], heavy fermion systems [27, 28], and topological materials [29–32]. However, unlike conventional ultrafast spectroscopy methods used to induce topological phase transitions, such as Floquet engineering [[33–35], coherent-phonon excitation [31], or lattice structure modification [32, 36], here we attempt to achieve topological transitions by tuning the magnetic configuration of EuAgAs. The metamagnetic nature of EuAgAs [20], coupled with the proximity of its Fermi level to Dirac points in all topological states [19, 20], makes it highly susceptible to external perturbations [18–20], including magnetic fields, strain, and pressure. These unique characteristics provide a direct and distinct route to controlling topological phase transitions, offering new insights beyond traditional topological manipulation techniques.

In this Letter, we report an ultrafast optical spectroscopy study of the response of the EuAgAs single crystal after it has been driven out of equilibrium with an ultrashort laser pulse. In particular, we observe an obvious change in the quasiparticle relaxation dynamics near $T_N$, indicating a modification of the electronic structure. Furthermore, the fluence dependence of the quasiparticle dynamics shows a different behavior above and below $T_N$. Below $T_N$, our high-fluence photoexcitation results suggest the possibility of photoinduced, non-thermal phase transitions. These results establish EuAgAs as a promising material for further investigation of the complex

* Corresponding author: peter.oppeneer@physics.uu.se
† Corresponding author: jqmeng@csu.edu.cn

relationship between topology and magnetism.

High-quality single crystals of EuAgAs were synthesized using Bi-flux methods, with a $T_N$ of approximately 12 K [20, 37] (see details in the Supplemental Material [38]). Ultrafast differential reflectivity $\Delta R/R$ measurements were carried out at a center wavelength of 800 nm ($\sim$ 1.55 eV) using a Yb-based femtosecond laser oscillator operating at 1 MHz with a pulse width of $\sim$ 35 fs. The pump and probe beams were focused onto the sample in a high vacuum environment ($< 10^{-6}$ Torr). The spot sizes were $\sim$140 and $\sim$60 $\mu$m for the pump and probe, respectively. The pump beam was $s$-polarized, and the probe beam was $p$-polarized. Further details of the experimental setup are available elsewhere [24].

Theoretical calculations show that the band structures of EuAgAs in the paramagnetic (PM) and AFM states are qualitatively similar, with changes in the Dirac points near the Fermi level [19, 20] [angle-resolved photoemission spectroscopy (ARPES) measurement also confirms these; see Fig. S4 in the Supplemental Material [38]]), whereas the band structure in the FM state deviates significantly due to exchange splitting [19, 20]. Theoretical calculations of the band structure suggest that the 1.55 eV photons used in the experiment primarily interact with the initial and final states of the Dirac cone. All these make ultrafast spectroscopy sensitive to changes in the electronic structure near the Fermi level in EuAgAs.

Figure 1(a) presents typical transient differential reflectivity $\Delta R/R$ as a function of delay time over a temperature range from 4 to 275 K at a low pump fluence of $\sim$13 $\mu$J/cm$^2$. Photoexcitation creates nonthermal electron and hole distributions that heat the electron gas, resulting in rapid $\Delta R/R$ changes. Thereafter, a rapid recovery of the change of photoinduced reflectivity occurred within $\sim$1 ps. Notably, the $\Delta R/R$ signal maintains a temperature-independent lineshape across the full range of measurements, but the amplitude of the signal exhibits strong temperature dependence, particularly around $T_N$. Below $T_N$, EuAgAs enters an AFM phase, and the change in $\Delta R/R$ amplitude suggests that the AFM ordering significantly modifies the low-energy electronic structure near $E_F$.

The two-dimensional (2D) pseudocolor map in Fig. 1(b) highlights this temperature dependence more clearly. In particular, around $T_N$, there is a noticeable suppression in the $\Delta R/R$ amplitude, which points to a phase transition driven by the onset of AFM order. This transition is accompanied by a substantial change in the electronic band structure, as the system evolves from a PM Dirac semimetal state to an AFM topological state. Additionally, a local minimum in amplitude is observed in the temperature range of 30-50 K, which closely corresponds to the magnetic fluctuations detected in transport measurements (see details in the Supplemental Material [38]).

A quantitative analysis of the quasiparticle dynamics was performed to examine its temperature-dependent behavior. The solid black lines in Fig. 1(a) indicate that, across all measured temperature ranges, the transient reflectivity over the

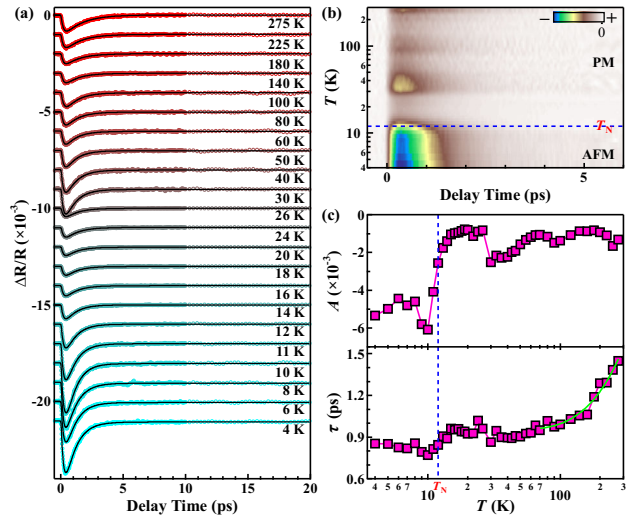

FIG. 1. **Temperature-dependent transient reflectivity in EuAgAs**. (a) Transient differential reflectivity $\Delta R/R$ as a function of delay time at various temperatures, ranging from 4 to 275 K, measured at a pump fluence of $\sim$13 $\mu$J/cm$^2$. The black solid lines are fits with Eq. (1). (b) 2D pseudocolor map showing $\Delta R/R$ as a function of both temperature and delay time. The logarithmic temperature scale highlights significant changes near $T_N$, indicating the influence of AFM ordering on the electronic structure. (c) Extracted amplitude $A$ (upper panel) and relaxation time $\tau$ (lower panel) as functions of temperature, showing clear anomalies around $T_N$. The green solid line represents a TTM fit.

measured time domain fits well with

$$\frac{R(t)}{R} = \frac{1}{\sqrt{2\pi}w}\exp\left(-\frac{t^2}{2w^2}\right)$$
$$\otimes \left\{\sum_{i=1}^{2} A_i \left[1 - \exp\left(-\frac{t}{\tau_{buildup}}\right)\right]\exp\left(-\frac{t}{\tau_i}\right) + C\right\}$$
(1)

where $A_i$ and $\tau_i$ represent the amplitude and relaxation time of the $i$th decay process, $\tau_{\text{buildup}}$ is the finite time required to build up the nonequilibrium quasiparticle population, $w$ is the incident pulse temporal duration, and $C$ is a constant offset. For the low-fluence regime, the transient reflectivity data can be well fitted with a single-exponential decay ($i = 1$).

Figure 1(c) summarizes the temperature dependence of amplitude $A_1$ and relaxation time $\tau_1$. The extracted parameters reveal clear anomalies at $T_N$, where both the amplitude and relaxation time undergo significant changes. In general, electron-electron scattering rapidly thermalizes the electron and hole distributions within tens of femtoseconds. Subsequently, hot carriers transfer their excess energy to the lattice system through electron-phonon ($e$-ph) scattering, which typically occurs on a picosecond timescale in semimetals [40–45]. At higher temperatures, EuAgAs has been suggested as a non-magnetic topological Dirac semimetal [19]. Similar to other Dirac semimetals [42–45], the ultrafast relaxation dynamics of hot Dirac fermionic quasiparticles in EuAgAs exhibit a comparable timescale ($\sim$1 ps), suggesting a dominant $e$-ph scattering mechanism. The solid green line in Fig.

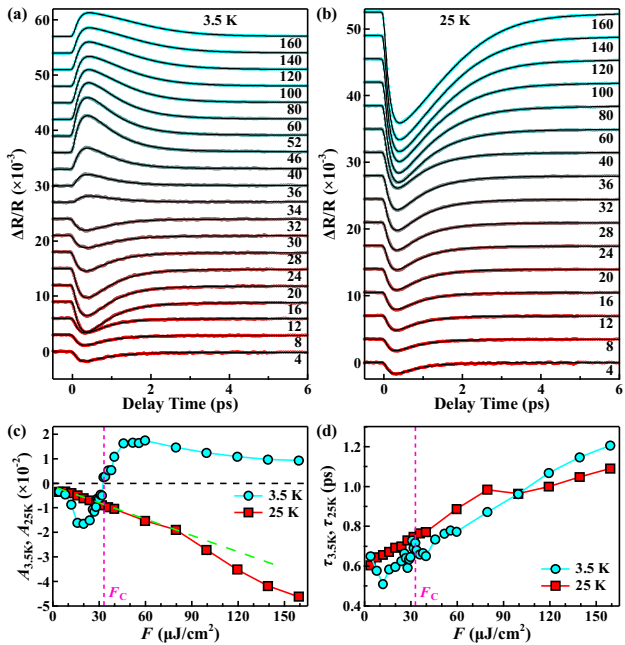

FIG. 2. **Fluence-dependence of transient reflectivity.** (a) and (b) Transient differential reflectivity $\Delta R/R$ as a function of delay time for various pump fluences, ranging from 4 to 160 $\mu$J/cm$^2$, measured at temperatures of 3.5 and 25 K, respectively. The black solid lines represent fits to Eq. (1) (c) and (d) Extracted amplitude $A$ and relaxation time $\tau$ of the initial decay, respectively, as functions of pump fluence $F$.

1(c) represents a fit using a two-temperature model (TTM) [40, 41], which describes the thermalization process between electrons and the lattice via $e$-ph scattering. The excellent fit across large temperature range indicates that $e$-ph scattering is the dominant relaxation mechanism in EuAgAs. However, it's important to note that in EuAgAs, electron-hole ($e$-$h$) recombination may also contribute to the fast decay process, as it does not require phonon mediation due to the overlapping conduction and valence bands at the same $k$ point.

Fluence-dependent measurements of transient reflectivity were employed to explore potential photoinduced phase transitions. Figure 2 shows the fluence-dependent transient reflectivity ($\Delta R/R$) measurements of EuAgAs, highlighting the distinct behavior of the material in both the AFM phase and the PM phase. These measurements were performed at 3.5 K (below $T_N$) and 25 K (above $T_N$), with the pump fluence varying from low to high, providing insight into how laser fluence influences the material's magnetic and electronic properties.

In Fig. 2(a), the transient reflectivity $\Delta R/R$ at 3.5 K exhibits a striking fluence-dependent change. As the pump fluence increases, a clear transition is observed at a critical fluence ($F_C \approx 33$ $\mu$J/cm$^2$), where the sign of $\Delta R/R$ switches from negative to positive. The change in sign suggests a modification of the electronic structure. In contrast, Fig. 2(b) shows the fluence dependence at 25 K, where the system is in the PM phase. Here, the transient reflectivity decreases monotonically with increasing fluence, and no sign reversal is observed. This suggests that in the PM phase, the system is less responsive to changes in fluence, and no complete phase transition is induced within the fluence range explored in this experiment.

To quantitatively analyze quasiparticle relaxation dynamics, we fitted the transient reflectivity data using Eq. (1). The transient reflectivity curves for the 3.5 K data fit well with a single-exponential decay across all pump fluences, as shown by the solid black lines in Fig. 2(a). For 25 K data, the transient reflectivity curves can be well fitted by Eq. (1) with a single-exponential decay for pump fluences below 80 $\mu$J/cm$^2$, as indicated by the solid black lines in Fig. 2(b). However, to achieve a better fit, a second component with positive amplitude must be included (see Fig. S5 in the Supplemental Material [38]). The emergence of a positive amplitude above $T_N$ signifies a different relaxation channel activated at high pump fluence, despite the overall negative transient reflectivity signal at 25 K. This suggests the possibility of localized light-induced modification of the electronic structure within the laser-irradiated region, similar to that observed at low temperatures.

Figures 2(c) and 2(d) summarize the extracted amplitude $A$ and relaxation time $\tau$ of the initial fast decay as functions of pump fluence $F$, respectively, for temperatures of 3.5 K and 25 K. At 3.5 K, as shown in Fig. 2(c), the amplitude ($A_{3.5\,K}$) initially decreases with increasing fluence, reaching a minimum around 15 $\mu$J/cm$^2$. As fluence approaches $F_C$, the amplitude crosses zero and becomes positive, indicating the onset of the FM phase. Beyond $F_C$, the amplitude increases sharply and saturates between $\sim$45 and 60 $\mu$J/cm$^2$. At 25 K, the amplitude ($A_{25\,K}$) shows a linear decrease with increasing fluence, without any abrupt transitions, further confirming that the PM state is less sensitive to fluence-driven phase changes. Similarly, Fig. 2(d) shows that the relaxation time at 3.5 K ($\tau_{3.5\,K}$) exhibits anomalies around $F_C$, gradually increasing as the fluence rises. This increase in $\tau_{3.5\,K}$ suggests that higher fluences lead to slower relaxation dynamics, likely due to the formation of a transient phase. In contrast, the relaxation time at 25 K ($\tau_{25\,K}$) follows a more linear trend, showing a steady increase with fluence, but without the anomalies seen at 3.5 K.

The fluence-dependent behavior of EuAgAs reveals a clear distinction between its response in the AFM and PM phases, providing strong evidence that EuAgAs is highly tunable via ultrafast optical excitation. The question arises: What accounts for the fluence dependence of such a large difference in transient reflectivity between below and above $T_N$? This stark distinction underscores the potential influence of distinct low-energy electron structures, such as magnetic topological mirror semimetals [19] or Dirac semimetals [20] in the AFM phase and Dirac semimetal in the PM phase [19], on carrier dynamics following laser excitation in EuAgAs. We consider three possible mechanisms to explain this fluence-dependent disparity:

(i) Laser-induced thermal effects: Intense laser pulses could substantially heat the sample surface, potentially inducing a phase transition from AFM to PM. However, this is unlikely to account for the sharp fluence threshold and abrupt sign reversal in reflectivity observed below $T_N$. The uniform negative $\Delta R/R$ response at lower fluences (Fig.1), regardless of whether the sample is in the AFM or PM phase, suggests that simple thermal effects alone do not explain the fluence-dependent behavior.

(ii) Phase-space [45] or band-filling [46] effects: High fluence excitation could lead to the saturation of available electronic states near the Dirac node, limiting further optical absorption due to Pauli blocking. This would result in a positive $\Delta R/R$ response. However, such effects should occur similarly in both the AFM and PM phases, independent of magnetic ordering. The absence of a similar response above $T_N$ suggests that phase-space filling is not the primary cause of the fluence-dependent behavior below $T_N$.

(iii) Photoinduced, nonthermal phase transitions, such as changes in magnetic order [47]: The most likely explanation is a photoinduced phase transition closely tied to the material's magnetic order. Given that 1.55 eV photons can effectively interact with Eu 4$f$ orbital electrons [30, 48] (see ARPES results in Fig. S3 in the Supplemental Material [38]) and the metamagnetic nature of EuAgAs, it is plausible that increasing fluence beyond $F_C$ alters the magnetic configuration [49, 50]. This, in turn, modifies the electronic structure. Based on previous theoretical [19, 20] and experimental [18, 20] work, we propose that this light-induced transition could involve a FM order, accompanied by a topological phase transition from a Dirac semimetal to a Weyl semimetal. This transition manifests as a fluence-dependent sign reversal in $\Delta R/R$, with the $F_C$ marking the onset of the FM phase. Above $T_N$, the system is already in the PM state, and while high fluence may perturb the magnetic configuration, it does not induce a full transition to the FM state. This results in a monotonic decrease in reflectivity without a sign reversal, but with the appearance of a positive amplitude.

To further test our hypothesis, a detailed temperature-dependent measurement was carried out with a higher fluence of $\sim$60 $\mu$J/cm$^2$. Figure 3(a) displays transient reflectivity $\Delta R/R$ at selected temperatures, while Fig. 3(b) presents the 2D pseudocolor map of $\Delta R/R$ as a function of pump-probe delay time and temperature. At low temperatures, the radiation region is already in the photoinduced transient phase. As the temperature increases, the transient reflectivity $\Delta R/R$ shifts from positive to negative at approximately 10.5 K, which is slightly below $T_N$. It can be found that the relaxation process at higher temperatures shows similarities to that observed at low fluence [Fig. 1(a)]: Within the magnetic fluctuation temperature region, complex temperature-dependent behavior occurs, such as the wave shape of $A_1(T)$ and $\tau_1(T)$, and the high-temperature PM phase exhibits a relaxation process resembling the $e$-ph scattering relaxation described by the TTM [solid green line in Fig. 3(c)]. It suggests that, at temperatures above $\sim$10.5 K, EuAgAs is dominated by the PM state. This also actually means that the transient state is difficult to be induced from the PM state by the laser, which is consistent with the conclusion of the fluence-dependent measurement results at 25 K. The high-fluence temperature-dependent experiment further substantiates the intimate correlation between the photoinduced phase transition and the AFM order.

Based on these observations, we propose a possible scenario for the photoinduced behavior in EuAgAs. In the AFM

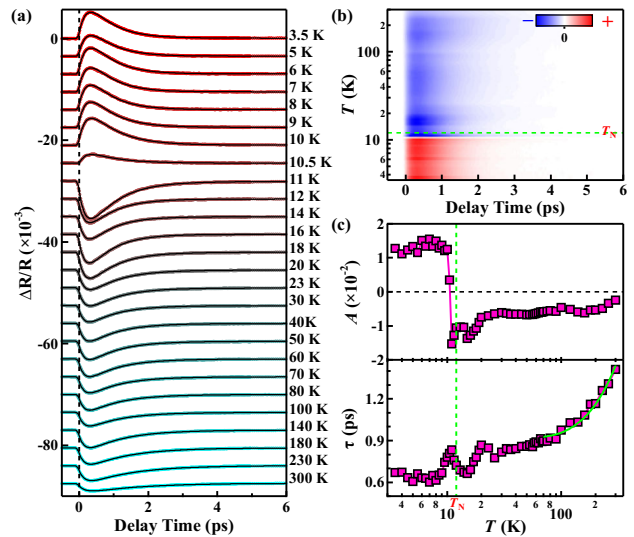

FIG. 3. **Evolution of ultrafast reflectivity with temperature in EuAgAs**. (a) $\Delta R/R$ as a function of delay time over a wide temperature range from 3.5 to 300 K, measured at a higher pump fluence ($\sim$60 $\mu$J/cm$^2$). The black solid lines are fits with Eq. (1). (b) 2D pseudocolor map of $\Delta R/R$ as a function of temperature and delay time. (c) Extracted amplitude $A$ (upper panel) and relaxation time $\tau$ (lower panel) as a function of temperature. The solid green line show the TTM fit.

phase, the system starts with a pair of Dirac cones at low pump fluences. Increasing the laser fluence triggers spin-flip transitions, leading to increased magnetization and eventually driving a phase transition to a FM Weyl semimetal state beyond a critical threshold ($F_C$). In the PM phase, the system similarly begins with a pair of Dirac cones at low fluences. While an increase in fluence may lead to a change in magnetization, a complete PM-to-FM phase transition is not observed within the studied fluence range. This suggests that additional factors, such as temperature or external magnetic fields, may be required to achieve a complete transition in the PM phase. To gain a more definitive understanding of the photoinduced phase and its dynamics, future studies should employ advanced techniques such as spin-polarized time- and angle-resolved photoemission spectroscopy, or time-resolved magneto-optical Kerr effect (TR-MOKE). These techniques can directly probe the electronic structure dynamics and magnetic order, providing crucial information to confirm the nature of the photoinduced phase and the underlying mechanisms driving the observed transitions.

This study uses ultrafast optical spectroscopy to investigate the nonequilibrium response of EuAgAs single crystals to ultrashort laser pulse excitation. Our findings reveal a strong correlation between magnetic configuration and electronic structure, evident from the shifts in spectral features under different conditions. The rapid response observed in the time-resolved data suggests that light can effectively induce transitions between different states. Our results show that EuAgAs, with its highly tunable magnetic structure, offers a platform for exploring light-induced topological phase transitions, possibly driven by changes in the magnetic order.

These findings underscore the potential of ultrashort optical pulses as tools for influencing and understanding the topological phase transitions, opening possibilities for advancements in spintronics, quantum computing, and energy-efficient information technologies.

This work was supported by National Key Research and Development Program of China (Grant No. 2022YFA1604204), the National Natural Science Foundation of China (Grant No. 12074436), the Science and Technology Innovation Program of Hunan province (Grant No. 2022RC3068), and Fundamental Research Funds for the Central Universities of Central South University (No. 1053320215412). We further acknowledge support from the K. and A. Wallenberg Foundation (Grants No. 2022.0079 and No. 2023.0336). This work is carried out in part using computing resources at the High Performance Computing Center of Central South University.

# Supplemental Material: Ultrafast photoinduced phase transition in the antiferromagnetic Dirac semimetal EuAgAs

Hao Liu,[1] Chen Zhang,[1] Qi-Yi Wu,[1] Yahui Jin,[2] Ziming Zhu,[2] Jiao-Jiao Song,[1] Sheng-Tao Cui,[3] Zhe Sun,[3] Honghong Wang,[4] Bo Chen,[1] Jun He,[1] Hai-Yun Liu,[5] Yu-Xia Duan,[1] Peter M. Oppeneer,[6,*] and Jian-Qiao Meng[1,†]

[1]*School of Physics, Central South University, Changsha, Hunan 410083, China*

[2]*Key Laboratory of Low-Dimensional Quantum Structures and Quantum Control of Ministry of Education, Department of Physics, Hunan Normal University, Changsha, Hunan 410081, China*

[3]*National Synchrotron Radiation Laboratory, University of Science and Technology of China, Hefei 230029, China*

[4]*Center for Quantum Materials and Superconductivity (CQMS) and Department of Physics, Sungkyunkwan University, Suwon 16419, South Korea*

[5]*Beijing Academy of Quantum Information Sciences, Beijing 100085, China*

[6]*Department of Physics and Astronomy, Uppsala University, Box 516, S-75120 Uppsala, Sweden*

E-Mail: peter.oppeneer@physics.uu.se

jqmeng@csu.edu.cn

**The file includes:**

1. Single Crystal Synthesis
2. XRD and EDS Characterization of EuAgAs Single Crystal
3. Resistivity, Magnetic Susceptibility and Magnetization
4. ARPES Results
5. Exponential Decay Fitting

1. **Single Crystal Synthesis**

Single crystals of EuAgAs were grown in Bi flux. Elemental metals with the ratio of Eu:Ag:As:Bi = 1:1:1:9 were placed in an alumina crucible and then sealed into an evacuated quartz tube. The mixture was heated to 1050 °C for 24 hours and slowly cooled to 700 °C at 2 °C/hour. Planar single crystals with typical dimensions of 4×4×2 mm$^3$ were extracted from the Bi flux using a high-speed centrifuge.

2. **XRD and EDS Characterization of EuAgAs Single Crystal**

Figure S1(a) displays the x-ray diffraction (XRD) pattern of the EuAgAs single crystal. All observed diffraction peaks can be indexed by (0, 0, $h$), with $h$ being even, indicating a pure single phase of the obtained single crystal. The chemical composition of the single crystal was characterized using a scanning electron microscope (SEM) equipped with an energy dispersive spectrometer (EDS). Figures S1(b) and S1(c) show the EDS results, confirming that the element ratio Eu:Ag:As is very close to 1:1:1.

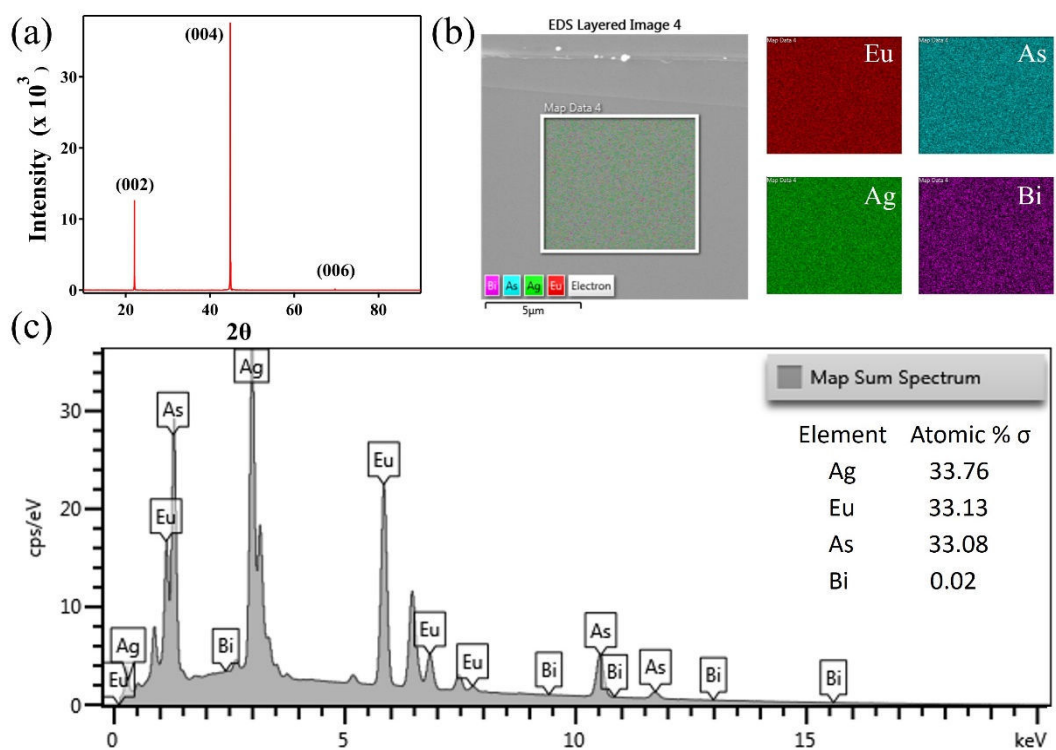

**Fig. S1**. XRD and EDS analysis. (a) XRD pattern. All peaks are indexed by (0, 0, $h$). (b) Surface morphology and elemental EDS map of the sample after the experiment discussed in the main text. Inset shows the area of EDS spectrum. (c) Averaged EDS spectrum of the region shown in (b). The atomic ratio of Eu:Ag:As is very close to 1:1:1 as expected.

## 3. Resistivity, Magnetic Susceptibility and Magnetization

A physical property measurement system (PPMS) was used to measure resistivity, magnetic susceptibility, and magnetization. Temperature-dependent resistivity [Fig. S2(a)] and zero field-cooled (ZFC) and field-cooled (FC) susceptibility ($\chi$) measurements for both the *c*-axis ($B \parallel c$) and the *ab*-plane ($B \parallel ab$) display a coincident peak at around 12 K [Fig. S2(b)], signifying a transition to antiferromagnetic (AFM) order at the Néel temperature ($T_N \approx 12$ K). The $\rho_{ab}(T)$ data [Fig. S2(a)] suggests metallic behavior well above $T_N$, transitioning to scattering-dominated behavior around $T_N$ due to critical fluctuations associated with the AFM ordering process. Below $T_N$, the decrease in $\rho_{ab}$ signifies the development of long-range AFM order. Magnetic anisotropy is evident from Fig. S2(c), where the magnetization $M(T)$ for $B \parallel ab$ exhibits a faster rise and saturation at a lower field compared to $B \parallel c$, consistent with previous reports [1]. The $\chi(T)$ curve for $B \parallel ab$ deviates from typical AFM behavior below $T_N$, exhibiting a non-monotonic temperature dependence with a rise after the initial drop. This, along with the observed bifurcation between ZFC and FC curves at low temperatures, suggests the presence of a weak ferromagnetic (FM) component and possible competition between AFM and FM interactions [1,2].

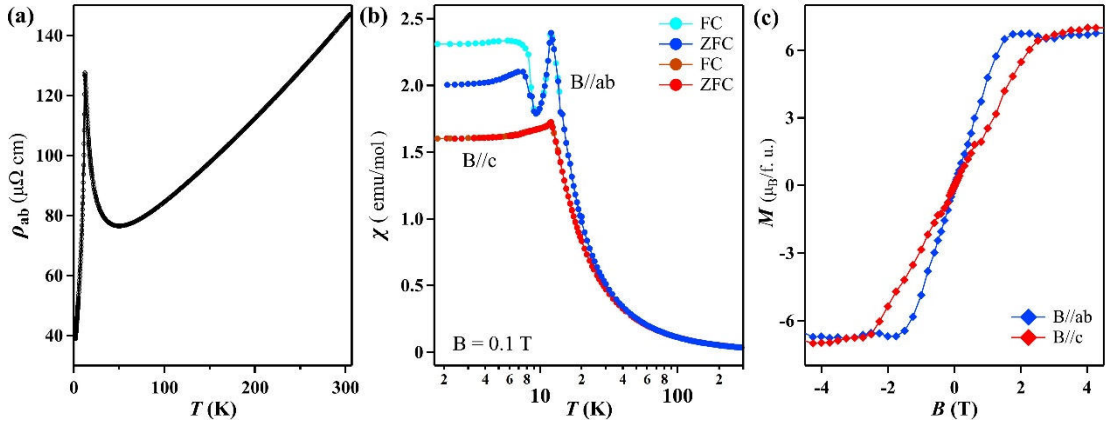

**Fig. S2**. Characterization of EuAgAs. (a) Temperature dependence of zero-field resistivity. (b) Magnetic susceptibility measured under zero-field-cooled (ZFC) and field-cooled (FC) conditions for $B \parallel c$ and $B \parallel ab$. (c) Magnetization ($M$) as a function of magnetic field ($B$) at 2 K for $B \parallel c$ and $B \parallel ab$ orientations.

## 4. ARPES Results

ARPES measurements were performed at the beamline 13U of the National Synchrotron Radiation Laboratory (NSRL, Hefei) equipped with a Scienta DA30 analyzer. All samples were cleaved *in situ* at low temperature along the (001) plane and measured under ultrahigh vacuum better than $6 \times 10^{-11}$ mbar.

Figure S3 presents ARPES maps of EuAgAs at 10 K, measured with 26 eV photons with a wide energy range. The Eu 4*f* orbital state is located about 1.5 eV below the Fermi level, similar to other Eu-based compounds like EuIn$_2$As$_2$ [3]. This suggests that 1.55 eV photons can effectively interact with Eu 4*f* orbital electrons, leading to the possibility of photon manipulation of the magnetic structure.

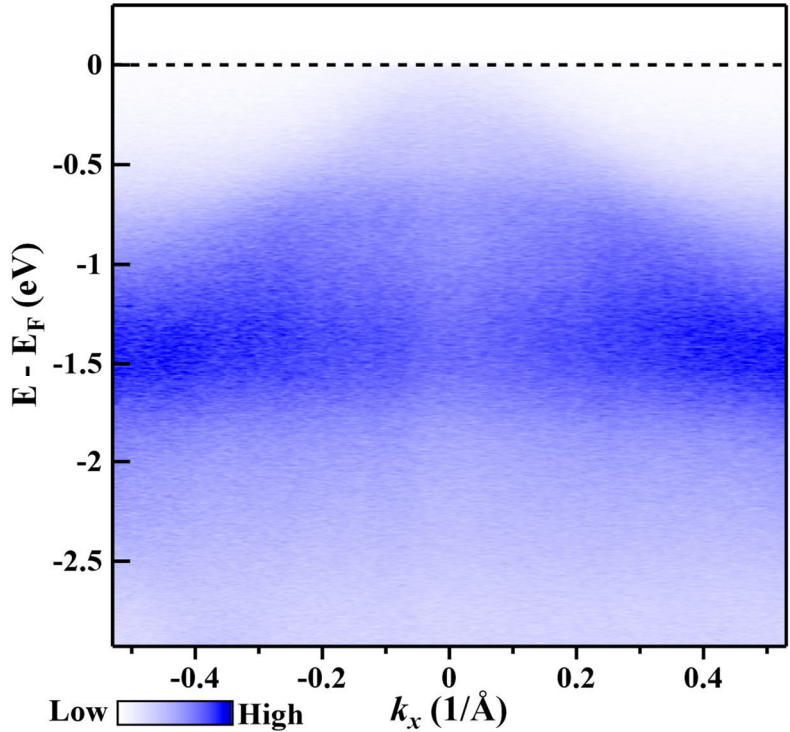

**Fig. S3**. ARPES maps of the EuAgAs band structure at 10 K, measured with 26 eV photons over a wide energy range.

Figures S4 (a) and (c) display the energy-momentum images of EuAgAs along the Γ'-K' direction at marked temperatures for 26 eV and 34 eV photon energies, respectively. Figures S4 (b) and (d) show the corresponding quasiparticle dispersions, obtained by applying a second derivative along the momentum direction to enhance the band structures while preserving the main features. As the temperature increases past $T_N$, the overall band structures remain qualitatively similar, consistent with theoretical calculations. Only a slight upward shift of the dispersions is observed, most clearly in Fig. S4 (d).

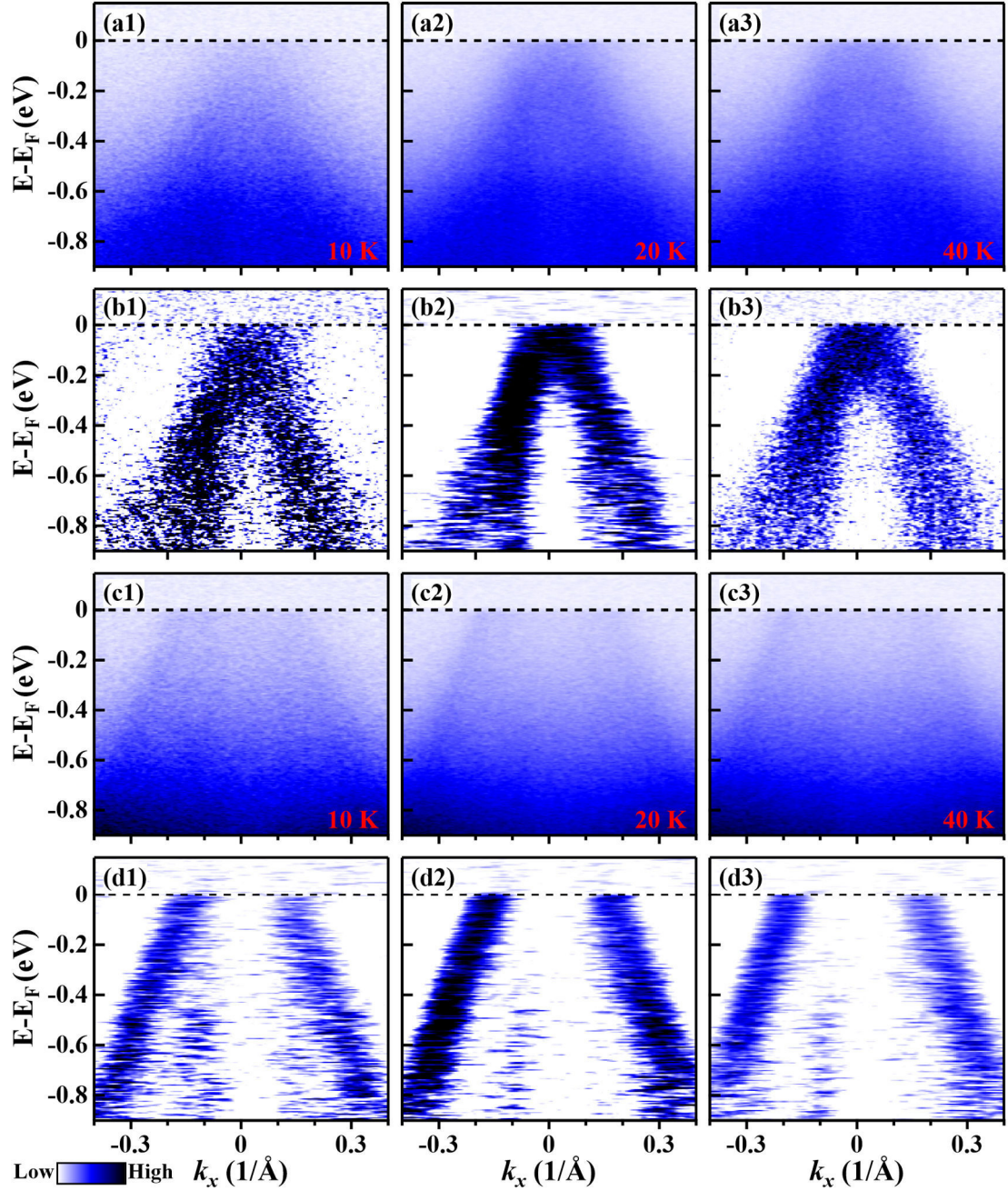

**Fig. S4**. ARPES maps of the EuAgAs band structure. (a1–a3) Band structure along the Γ'-K' direction at different temperatures measured with 26 eV photons. (b1–b3) Second-derivative images of (a1–a3) to highlight weak bands. (c1–c3) Band structure along the Γ'-K' direction at different temperatures measured with 34 eV photons. (d1–d3) Second-derivative images of (c1–c3) to highlight weak bands.

## 5. Exponential Decay Fitting

Fig. S5 presents the results of fitting the transient reflectivity at a pump fluence of 160 µJ/cm$^2$ and a temperature of 25 K using single- and dual-exponential decay functions. The solid black line in the upper panel shows that a single-exponential decay function does not accurately fit the transient reflectivity curves. In contrast, a dual-exponential function provides a satisfactory fit, as shown in the lower panel. Therefore, we used the dual-exponential function for fitting the data at $F \geq 80$ µJ/cm$^2$.

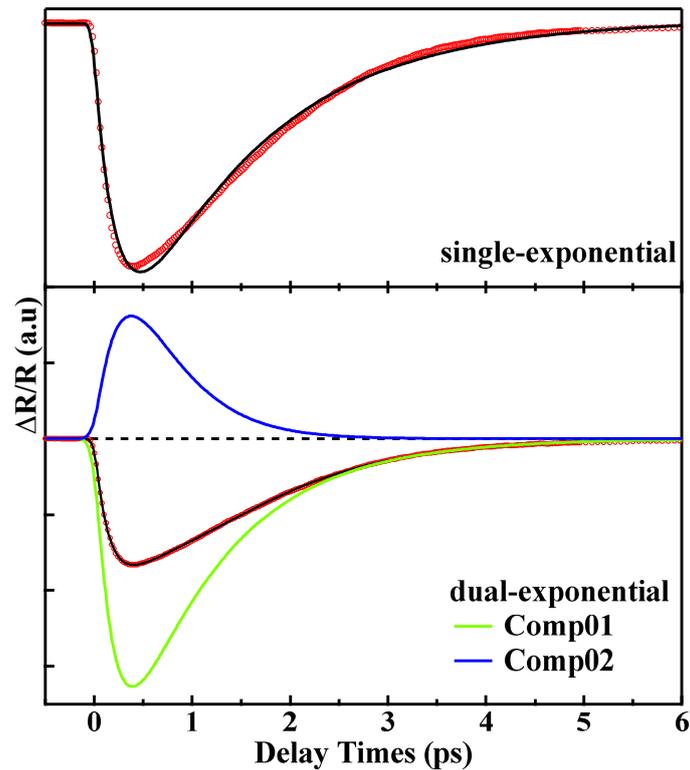

**Fig. S5**. Fitting of the transient reflectivity signal at a pump fluence of 160 µJ/cm$^2$ and a temperature of 25 K using single-exponential decay (upper panel) and dual-exponential decay (lower panel) functions.